

\magnification \magstep 1
\vsize=24 true cm
\hsize=15 true cm
\baselineskip= 0.6 true cm
\font\big=cmr10 scaled\magstep 1
\font\bigre=cmbx10 scaled\magstep 1
\font\bigra=cmbx10 scaled\magstep 2

\vglue 2 true cm
\centerline{\bigra Traveling Waves in a Fluid Layer}
\vskip 0.5 true cm
\centerline{\bigra Submitted to a Horizontal Temperature Gradient}
\vskip 1 true cm

\centerline {\big F. Daviaud and J.M. Vince}

\vskip 1 true cm

\centerline{\sl Service de Physique de l'Etat Condens\'e,
CEA - Centre d'Etudes de Saclay}
\vskip 0.3 true cm
\centerline{\sl F-91191 Gif-sur-Yvette Cedex, France}

\vskip 2 true cm

We report experimental observations of traveling waves in a
pure fluid with a free surface situated
in a long container submitted to a horizontal
temperature gradient perpendicular to its large extension.
Above a critical value of the gradient and depending on the height
of liquid $h$, a source of waves is created in the container  for
small value of $h$, while  the system exhibits stationary patterns
for larger values of $h$.
The spatio-temporal properties of the waves are studied
and compared to theoretical predictions.

\vfill\eject

While the nature of turbulence in fluids is still today an
open question, the evolution towards turbulent states has
received much attention  in the last decade and  important results
have been obtained in weakly confined systems.
In this frame, the study of one-or two- dimensional (1D or 2D) systems
in numerical models and in experiments has shown that a
periodic cellular state may destabilize when a control parameter
is varied and becomes turbulent through different scenarios.
A transition to turbulence via spatio-temporal intermittency has
been observed in models of phase equations [1] and coupled map lattices
[2] as well as in 1D Rayleigh-B\'enard convection [3], in the
Printer's instability [4] or in capillary ripples [5].
A similar regime is also observed in circular [6]
  or plane Couette [7] flows.
Spatio-temporal chaos can also appear by an increase of the
number of defects as it is seen in Ginzburg Landau equations [8]
or in convection in binary fluid mixtures [9]. Much work has been
devoted to this last system in which the first state observed consists of
traveling waves which arise from a subcritical Hopf bifurcation.
This system displays a rich variety of dynamical states, such as confined
states or defects and  spatio-temporal chaos near onset
in  1D [10] and 2D systems [11].
More recently, 1D propagative patterns have also been observed in
a new experiment involving a hot wire below the free
surface of a liquid. This dynamic state is qualitatively consistent
with theories based on Ginzburg-Landau equations [12].
 However, the physical mechanisms responsible for this instability
 remain unknown as well as the exact  role of defects.

In this letter, we report new experimental results concerning different
dynamical regimes, including traveling waves,
obtained in a very simple configuration.
A pure fluid with a free surface is situated in a long and narrow
container and is submitted to a horizontal temperature gradient
perpendicular to its large extension.
Many experiments have been devoted to this problem,
 but always in containers with a small
extension perpendicular to the gradient [13]. In this
case, the propagating modes were never observed, whereas they can
develop in our geometry. This problem has also received increasing
attention in the past decade, because of microgravity facilities,
in methods of float-zone crystal-growth with cylindrical geometries [14].
It was theoretically studied,
 in the case where gravity is ignored [15,16,17].
Our system thus appears to be a good tool
to study both the mechanisms which drive this instability and
the dynamics of nonlinear 1D traveling waves.

\vskip 0.5 true cm

The apparatus consists of a rectangular container 20 cm long
and 1 cm large (see Fig.1). The long vertical walls are made of copper
and can be thermally regulated by circulating water. The container
is closed by two small plexiglass walls.
The lower boundary consists of a glass plate to allow visual
observation while a plexiglass plate is inserted a few
millimiters above the surface of the fluid to avoid evaporation
problems. The container is filled with silicon oil
(viscosity $\nu=0.0065$ Stokes) of Prandtl number $P=7$
up to a height $h$ that is measured with a precision of 0.05 mm.
The horizontal temperature gradient $\Delta T$ is imposed
between  the two copper walls and is measured using thermocouples.
This gradient is regulated with a stability of $10^{-2}$ K.
\par
The patterns are visualized by shadowgraphic imaging. A
parallel light beam crosses vertically the container from top
to bottom and forms a horizontal picture on a screen,
both due to  surface deformations at the interface oil-air
and to temperature gradients in the fluid (see Fig.2).
The spatio-temporal evolution of the structures is recorded
using a video camera and the images are digitized along a
line of 512 pixels perpendicular to the gradient.

\vskip 0.5 true cm

The two parameters which  control our experimental system
are $h$, the height of Si oil in the cell, and  $\Delta T$,
the horizontal temperature difference between the two walls.
We have observed that, for each $h$ value,
when $\Delta T$ is increased from zero to a threshold value
 $\Delta T_c$, a longitudinal (parallel to the gradient)
pattern appears. But depending on the height of liquid,
two different regimes take place. When $\Delta T \ge \Delta T_c$,
for  small $h$ values (from 0 to 2.8 mm),
the system exhibits longitudinal waves propagating
  perpendicularly to the gradient
 (see Fig.2) while, for  larger $h$ values (from 2.8 mm to 10 mm),
stationary \lq\lq rolls\rq\rq  with axis parallel
 to the gradient are observed.
 Figure 3 shows the dependence of $\Delta T_c$ on $h$ for
 these two kinds of structures. One can notice that the
value of height $h=2.8$ mm for which the two domains meet each other
corresponds to a ratio between the Marangoni and the Rayleigh
number $W=Ma/Ra\simeq 1$ with
$Ma=-(\partial\sigma/\partial T)\Delta T h /\rho_0 \kappa \nu$ and
$Ra=g\alpha\Delta T h^3/\kappa \nu $,
where $g$ is the gravitational acceleration, $\alpha$ the thermal
expansion coefficient, $\kappa$ the thermal diffusivity, $\rho_0$
the density of the fluid, $\nu$ the kinematic viscosity and $\sigma$
the surface tension. The traveling waves domain corresponds
to $W>1$, which means that surface tension effects are dominant,
while in the stationary rolls domain ($W<1$), buoyancy is
preponderant.

When studying the regime of propagative waves in more details,
one can observe that the structures are not exactly perpendicular
to the large side of the container but exhibit an angle
$\psi \simeq 80^{\circ}$. This angle does not seem to
depend on $\Delta T$ or on $h$.
When $\Delta T$  is increased above $\Delta T_c$,
a source of waves is created with a dimension which depends
on $\varepsilon=(\Delta T - \Delta T_c)/ \Delta T_c$.
 The size of the source is very large near the threshold
and decreases in length as $\varepsilon$ is increased.

Space-time diagrams have been performed to study the dynamic of
the waves (see Fig. 4). With such diagrams,
 the wavelength and the period of these waves can easily be measured.
First of all, spatial and temporal Fourier spectra reveal
that the system of waves has a unique
 wavelength and is monoperiodic at threshold.
For $h$ values between 0 and 2.5 mm, the period at threshold
 $T_c$ increases with $h$ (cf. Figure 5).
In the same range of $h$, the $\lambda_{c}/h$ variation
versus $h$ is  no-monotonous, with a maximum for $h=1.4$ mm.
These two behaviours can be compared to those
obtained   when  a hot wire
is  used to destabilize a fluid layer with a free surface [12].
The physical context is somewhat different but
the behaviour of the two sets of curves is very similar,
allowing us to think that the physical mechanism which
drives the instability in the hot wire experiment could
 have the same origin as in our configuration.

We have also performed some measurements for  $\Delta T$
larger than $\Delta T_c$ at different $h$ values.
The curves giving the period $T$
and the phase velocity $v_{\varphi}$ versus $\varepsilon$
are respectively displayed in Fig.6a and Fig.6b.
The  figures show  that $T$
decreases and $v_{\varphi}$ increases when $\varepsilon$
is increased. One can notice that, for
small $h$ values, the  values of $T$ and $v_{\varphi}$
are clearly different from these obtained for the middle
range of $h$ values. This fact  shows how this phenomenon is
sensitive to the surface tension effects which are more
important as $h$ is small.
Finally, when $\varepsilon$ is further increased, the
system of traveling waves exhibit phase instabilities
leading to space time dislocations.

\vskip 0.3 true cm
For the stationary rolls, the experiment shows that they are
strictly perpendicular to the large side of the cell,
contrary to traveling waves. The wavelength associated to these
 rolls appears to be proportional to the height of liquid $h$.
Then, when $\varepsilon$ is increased, the rolls begin to
oscillate under the form of an optical mode,
 and, for larger $\varepsilon$, the pattern is destabilized and
gives place to spatio-temporal chaos.
These dynamical regimes are reminiscent of what
is observed in  Rayleigh-Benard convection in narrow gap geometries [3].

\vskip 0.5 true cm

Our observations raise several interesting issues concerning
both the hydrodynamical and the phenomenological aspects.
First of all, thermocapillarity appears to be the main physical
mechanism driving the instability observed in our experiment
for small values of $h$, while buoyancy is preponderant for
larger values of $h$.
 It is known that, due to thermocapillary effects, fluid motion
can develop when a temperature gradient is imposed on a thin
fluid layer with a free surface [16].
 Although we have not yet recorded velocity profiles, by seeding the flow
with particles and illuminating it with a laser sheet,
we have seen that the horizontal velocity
profile  reveals a shear flow near the surface
and a return flow at the bottom, the convection being in the
form of a monocellular cell (cf. [13]). Smith and Davis have
shown that this dynamic state can destabilize under the
form of two types of thermoconvective instabilities:
stationary longitudinal rolls and traveling hydrothermal waves [15].
Some phenomena, such as temperature oscillations, had been
previously observed in floating zones experiments [14],
but, to our knowledge, our observations provide the first clear
evidence of these instabilities.
However, our experimental results share some differences
with the results obtained by Smith and Davis.
These authors consider the destabilization of two basic flow profiles:
 the linear flow
state (LFS) and the return flow state (RFS). Stationary rolls
are not stable in RFS but only in LFS. Moreover, oblique traveling
waves are not predicted in fluids with $P>1$ in LFS
while they do exist in RFS, but with an angle $\psi  \simeq 20^{\circ}$
instead of $80^{\circ}$ as observed in the experiments.
\par
These differences could be explained by the absence of gravity
in their model. More recently, the stability of
convective motion in a  differentially heated cavity
has been studied, in the presence of gravity [18,19,20].
 However, certainly due to different boundary
conditions, our experimental observations differ from their results.
A more realistic model including both surface tension and gravity
together with appropriate boundary conditions is at the moment studied
and the first results are promising [21].

As for the phenomenological aspect, our configuration reveals to be
 a good system for exploring nonlinear 1D traveling waves
and their destabilization.
The spatio-temporal properties of the waves can be studied
in the frame of two coupled amplitude equations, that could be
derived from the physical model [21]. As a matter of fact,
they  have  general properties that do not depend
on the exact experimental system. In our
experiement, the presence of a source
stabilizes the right- and left-going waves with a definite
wavelength  and  near threshold, the width of
the source increases when decreasing $\varepsilon$, as it is
observed in the hot wire experiment [12] and in recent
numerical simulations [22]. Our experiment  also shows
 similarities with some dynamical regimes
(such as confined states with a stable sink of rolls) oberved in
binary convection in an annulus [9,10].
We are performing experiments in this
direction, using an annular configuration with a radial temperature
gradient.

\vskip 0.3 true cm
 We wish to thank M.Dubois, J.Lega and C.Normand for
stimulating discussions and M.Labouise, P. Hede and B. Ozenda
for their technical assistance.

\vfill\eject

\noindent{\bigre References}

\vskip 1 true  cm

\item{[1]} H.Chat\'e and P.Manneville
Phys.Rev.Lett. {\bf 58}, 112 (1987)

\item{[2]} K.Kaneko
Prog.Theor.Phys. {\bf 74}, 1033 (1985)
\item{} H.Chat\'e and P.Manneville,
Physica D {\bf 32}, 409 (1988)

\item{[3]} S.Ciliberto and P.Bigazzi, Phys.Rev.Lett. {\bf 60}, 286 (1988)
\item{} F.Daviaud, M.Dubois and P.Berg\'e,
Europhys.Lett. {\bf 9}, 441 (1989)

\item{[4]} M.Rabaud, S.Michalland and Y.Couder,
Phys.Rev.Lett. {\bf 64}, 184 (1990)

\item{[5]} N.B.Tufillaro, R.Ramshankar and J.P.Gollub,
Phys.Rev.Lett. {\bf 62}, 422 (1989)

\item{[6]} C.D.Andereck, S.S.Liu and H.L.Swinney,
J.Fluid Mech. {\bf 164}, 155 (1986)

\item{[7]} F.Daviaud, J.Hegseth and P.Berg\'e,
Phys.Rev.Lett. {\bf 69}, 2511 (1992)

\item{[8]} P.Coullet and J.Lega,
 Europhys.Lett. {\bf 7}, 511 (1988)

\item{[9]} see e.g. D.Bensimon, P.Kolodner, C.M.Surko, H.Williams
and V.Croquette,
J. Fluid Mech. {\bf 217}, 441 (1990).

\item{[10]} P.Kolodner, J.A.Glazier and H.Williams,
Phys.Rev.Lett. {\bf 65}, 1579 (1990)
\item{} P.Kolodner,
Phys.Rev. A {\bf 46}, 6431 (1992)

\item {[11]} V.Steinberg, E.Moses and J.Fineberg,
Nucl.Phys. {\bf B2}, 109 (1987)

\item{[12]} J.M. Vince and M.Dubois,
Europhys.Lett. {\bf 20}, 505 (1992)

\item{[13]} see e.g. D.Villers and J.K.Platten,
J. Fluid Mech. {\bf 234}, 487 (1992).

\item{[14]} F.Preissier, D.Schwabe and A.Scharmann,
J. Fluid Mech. {\bf 126}, 545 (1983)
\item{} J.J.Xu and S.H.Davis,
Phys.Fluids {\bf 27}, 1102 (1984)

\item{[15]} M.K.Smith and S.H.Davis,
J. Fluid Mech. {\bf 132}, 119 (1983).

\item{[16]} S.H.Davis,
Ann.Rev.Fluid Mech. {\bf 19}, 403 (1987).

\item{[17]} M.K.Smith,
J. Fluid Mech. {\bf 194}, 391 (1988).

\item{[18]} P.Laure and B.Roux,
J. Cristal Growth {\bf 97}, 226 (1989)

\item{[19]} B.M.Carpenter and G.M.Homsy,
J. Fluid Mech. {\bf 207}, 121 (1989).

\item{[20]} G.Z. Gershuni, P.Laure, V.M.Myznikov, B.Roux,
E.M.Zhukhovitsky,
preprint submitted to Europ.J.Mech. B (Fluids)

\item{[21]} C.Normand and J.Lega,
private communication

\item{[22]} F.Plaza, J.M.Vince and M.Dubois,
 private communication

\vfill\eject
\noindent {\bigre Figure Captions}
\vskip 1 true cm
\noindent{\bf Figure 1:} Schematic drawing of the experimental apparatus.
\vskip 0.4 true cm

\noindent{\bf Figure 2:} Shadowgraphic image of a traveling wave
pattern in an horizontal plane. The top (respectively the bottom)
of the picture corresponds to the hot (repectively the cold) plate.
\vskip 0.4 true cm

\noindent{\bf Figure 3:} Critical temperature difference $\Delta T_c$
vs. the height of liquid $h$.
\vskip 0.4 true cm

\noindent{\bf Figure 4:} Spatio-temporal evolution showing
a left- and a right going wave emitted from a source and
corresponding to the pattern shown in Figure 2.
 The total acquisition time is
25 sec and the spatial extension is 20 cm.
\vskip 0.4 true cm

\noindent{\bf Figure 5:} Evolution of the period $T_c$
 with the height of liquid $h$.
\vskip 0.4 true cm

\noindent{\bf Figure 6:} Evolution of the period $T$ (a)
and of the velocity  $v_{\varphi}$
(b) of the waves with the reduced temperature
difference $\varepsilon$.

\end